\documentclass[twocolumn,showpacs,preprintnumbers,amsmath,amssymb]{revtex4}
\usepackage{graphicx}
\usepackage{dcolumn}
\usepackage{bm}
\begin{document}

\title{Spin transitions in an incompressible liquid Coulomb coupled to a
quantum dot}
\author{V.M. Apalkov$^{1,2}$ and Tapash Chakraborty$^2$}
\affiliation{$^1$Department of Physics and Astronomy, Georgia State University,
Atlanta, Georgia 30303, USA \\
$^2$Department of Physics and Astronomy,
The University of Manitoba, 223 Allen Building, Winnipeg,
Canada R3T 2N2} 
\author{C. Sch\"uller}
\affiliation{
Institut f\"ur Experimentelle und Angewandte Physik,
Universit\"at Regensburg, D-93040 Regensburg, Germany}
\date{\today}

\begin{abstract}
We report on our investigation of the low-lying energy spectra and charge density
of a two-dimensional quantum Hall liquid at $\nu=\frac25$ that is Coulomb coupled 
to a quantum dot. The dot contains a hole and two/three electrons. We found that
any external perturbation (caused by the close proximity of the quantum dot)
locally changes the spin polarization of the incompressible liquid. The effect 
depends crucially on the separation distance of the quantum dot from the electron 
plane. Electron density distribution in the quantum Hall layer indicates 
creation of a quasihole that is localized by the close proximity of the quantum dot.
Manifestation of this effect in the photoluminescence spectroscopy is also
discussed.
\end{abstract}

\maketitle

The two-dimensional electron gas (2DEG) in an intense external perpendicular
magnetic field and at very low temperatures has proven to be a remarkable
system to explore various quantum phase transitions. Immediately after the
discovery of the fractional quantum Hall effect \cite{fqhe_expt} and the
subsequent seminal theoretical work by Laughlin \cite{laughlin} who introduced 
the idea of an incompressible quantum liquid to explain the effect, it became 
clear that there are other interesting transitions, for example, the spin 
transitions at a few special filling factors, that are 
entirely driven by the electron correlation. Spin unpolarized ground
states \cite{halperin,spin} and excitations \cite{spin_quasihole}
were proposed and intensely investigated theoretically 
\cite{spin-reversed,qhbook} and eventually confirmed in experiments
by various groups \cite{activated}. It is now widely accepted that the
$\nu=\frac25$ filling factor is spin {\it unpolarized} in its ground state
and has spin-reversed quasiholes as low-energy excitations \cite{qhbook}.
This is in contrast to the other fundamental filling factor $\nu=\frac13$
which has a fully spin-polarized ground state and spin-polarized
fractionally-charged quasiparticles and quasiholes \cite{laughlin,qhbook}.

The $\nu=\frac13$ fractionally-charged quasihole was probed recently 
in a very interesting experiment that involved photoluminescence (PL) 
spectroscopy of a two-dimensional electron gas confined in a quantum 
well subjected to an external magnetic field. In that experiment,
electron-hole pairs are excited with a single laser line, and 
afterwards, electrons and holes relax and form excitons. Studying the 
luminescence lines (when the electrons and holes recombine) as one 
sweeps the magnetic field, one can determine various physical properties
of the system in a magnetic field. The physics in this situation is 
seemingly well established. When the lowest Landau level is completely 
occupied the observed PL lines show a nearly linear  magnetic field 
dispersion that can be attributed to recombination of electrons from 
the occupied Landau levels with photocreated holes in the valence band. 
However, interesting things were found to happen at the intermediate density 
($0.9\times10^{11}$\ cm$^2$) and a moderate electron mobility. Here, 
Sch\"uller et al. \cite{schuller} observed that, near the 1/3 fractional quantum 
Hall region the PL lines exhibit a {\it large reduction} in energy  ($\sim$ 2 meV) 
at a very low temperature (T=0.1 K) compared to those at a much higher 
temperature (T=1.8 K) where the lines are as expected, linear. Further, 
a puzzling observation associated with the anomaly was that a very small 
thermal energy ($\ll 2$ meV) is sufficient to destroy the anomaly.

The anomaly was resolved by a theoretical model 
\cite{schuller,qhconfine} where we considered a situation in which 
the excitons are localized but they are in close proximity to a 
$\nu=\frac13$ incompressible liquid. In this model, localized 
exciton are represented by quantum dots (the nanometer-scale boxes for 
confining electrons and holes) \cite{chakmak} and the observed anomaly 
is related to the properties of the combined dot-plus-liquid system 
(a {\it liquid-qd complex}) where the quantum dot (QD) contains a charge-neutral 
exciton plus an extra electron. The close proximity of the dot perturbs 
the incompressible electron liquid and creates fractionally-charged 
quasiholes in the liquid. Interestingly, the calculated energy to create 
those quasiholes turned out to be similar to the tiny energy that was 
required to destroy the observed anomaly. This result explained the observed puzzle 
described above: application of a small thermal energy closes the quasihole 
energy gap and therefore the incompressibility of the liquid, the prime
reason for the existence of quasiholes, disappears and so does the anomaly
\cite{schuller}. The experiment and its theoretical explanation 
clearly demonstrated that exciton spectroscopy as opposed to 
transport measurements might provide an important route to explore the 
properties of the incompressible states. 

A question that naturally arises then is, what happens when the two-dimensional
electron liquid is in the $\nu=\frac25$ state where the ground state, at low magnetic 
fields, is expected to be spin unpolarized \cite{qhbook}, rather than in the
$\nu=\frac13$ state where the ground state is fully spin polarized. It is not 
immediately obvious how that state would evolve when a charged quantum 
dot containing one or two electrons plus an exciton is Coulomb 
coupled to that quantum Hall state. In this paper, we report on our investigation 
of the spin transitions in such a liquid-qd complex.

\begin{figure}
\begin{center}\includegraphics[width=9cm]{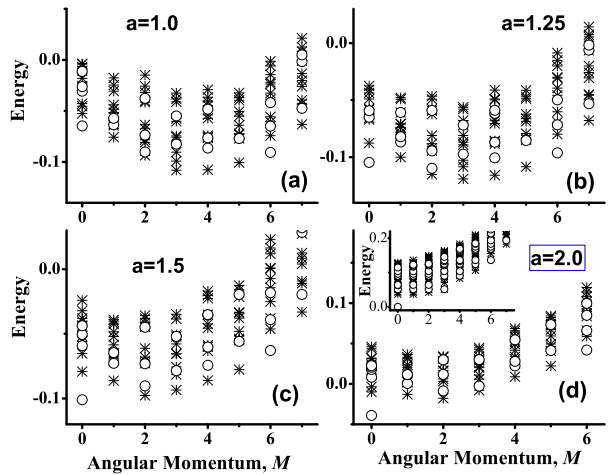}\end{center}
\vspace*{-1cm}
\caption{Energy (in units of Coulomb energy) versus the azimuthal
rotational quantum number of a Coulomb-coupled liquid-QD complex
at $\nu=\frac25$. The quantum dot contains two electrons and a hole  
and is separated from the 2DEG by $a=1.0 - 2.0\, l_0$,
In this figure, $\circ$ represents the energies of a 
spin unpolarized state, and the $\ast$ corresponds to the spin 
partially polarized state energy.}
\end{figure}

Details for calculation of the energy spectrum and
charge densities for the liquid-qd complex are available in 
Refs.~\cite{schuller,qhconfine}. Electrons and the hole in the QD are
confined by a parabolic potential $V_{\rm conf}(x,y)=\frac12m^*\omega_0^2
(x^2+y^2)$, where $\omega_0$ is the confinement potential strength. In 
our calculations the size of the quantum dot, $l_{\rm dot}=\left(\hbar/m^*
\omega_0\right)^\frac12$, was equal to 15 nm. We take into account
only eight states in the QD, which can be occupied either by electrons or the 
hole. Electrons and the hole in the dot are treated as spinless particles.
To model the electron layer we use the spherical geometry \cite{impure} with 
six electrons and twelve flux quanta. This corresponds to the electron filling 
factor $\nu=\frac25$. We consider the Coulomb interaction between the electrons
and the hole in the dot and the electron layer separated by a distance $a$.
For a charge neutral quantum dot containing one electron and one hole, 
the effect of the dot on the incompressible liquid is small. In this 
case we have an almost decoupled system of the 2D liquid and the quantum dot, 
similar to the case of $\nu =1/3$ observed earlier \cite{schuller}. The new 
features that are specific for $\nu =2/5$ appear when the quantum dot becomes 
negatively charged. 

In Fig.~1, the energy spectrum (in units of Coulomb energy, $e^2/\epsilon l_0$,
where $\epsilon$ is the background dielectric constant and $l_0^2=\hbar c/eB$
is the magnetic length) of the system with two electons and a hole in the 
quantum dot is shown. The open circles correspond to unpolarized 
states of the incompressible liquid, while $\ast$ describe the spin 
partially polarized states. In our case the partially polarized states 
correspond to four electrons with spin-down and two electrons with spin-up. The 
ground state of the unperturbed $\nu=2/5$ liquid has zero angular momentum and 
is spin unpolarized \cite{spin,qhbook}. When we introduce the interaction between 
the 2D layer and the quantum dot, the spin polarization of electrons in the electron layer 
crucially depends on the separation between the 2D layer and the quantum dot, i.e. 
on the strength of interaction between electrons in the layer and electrons and 
hole in the dot. For a small separation, $a=1.0\, l_0$ and $1.25\, l_0$, the ground state is 
spin partially polarized and has a finite angular momentum, $M=3$ (see Figs.~1a and 1b). 
These data illustrates that any external perturbation of the $2/5$-liquid changes 
locally the spin polarization of the incompressible liquid. This is in correpondence
with the energy spectrum of the unperturbed (without the presence of a dot)
$2/5$ liquid (shown as inset in Fig.~1d), where the lowest excitation of the 2D liquid 
is spin partially polarized. For large separation $(a\geq 1.5 l_0)$, the ground state 
is again spin unpolarized at $M=0$, as in the unperturbed case.

\begin{figure}
\begin{center}\includegraphics[width=9cm]{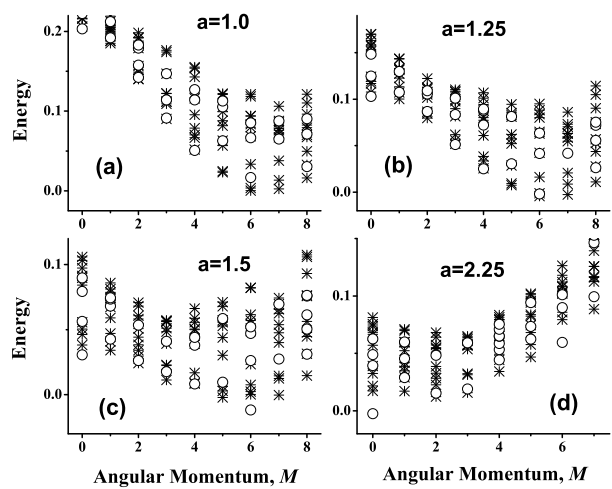}\end{center}
\vspace*{-1cm}
\caption{Same as in Fig.~1, but here the quantum dot contains three electrons 
and a hole  and is separated from the 2DEG by $a=1.0 - 2.0\, l_0$. The sysmbols are 
described in Fig.~1.}
\end{figure}

Figure~2 displays the energy spectra of the system containing three electrons 
and one hole in the dot. In this case the quantum dot effectively becomes doubly charged
(negative), which should increase the interaction strength between the quantum dot 
and the electron liquid. This increase can be clearly seen in Fig.~2. When the separation 
between the quantum dot and the layer is small ($a=1.0\, l_0$) the ground state is spin 
partially polarized and has a large angular momentum, $M=6$. Let us now compare these
results with those in Fig.~1a, where there are only two electrons in the dot and the angular 
momentum of the ground state is equal to three. The increase of the angular momentum of 
the ground state illustrates the effect of stronger interaction between the quantum dot 
and electrons in the 2D layer. When the separation between the 2D layer and quantum dot is 
increased, there appears a new feature which is not present in Fig.~1, namely, the 
ground state which in some region of $a$ becomes unpolarized at finite
angular momentum (Fig.~2c). Therefore for three electrons in the 
quantum dot we have a competition between the unpolarized and partilally 
spin polarized states at finite angular momentum, $M=6$. When the 
separation $a$ is further increased the ground state at first becomes 
partially polarized at a finite angular momentum, $M=2$, and then 
it becomes unpolarized at the zero angular momentum. This occurs at 
separations larger than in the case of two electrons in the dot
(Fig.~1). In Fig.~2d the separation between the quantum dot and the layer 
is $a=2.25\, l_0$ and the energy spectrum is similar to that of the unperturbed 
$2/5$ liquid. 

\begin{figure}
\begin{center}\includegraphics[width=9cm]{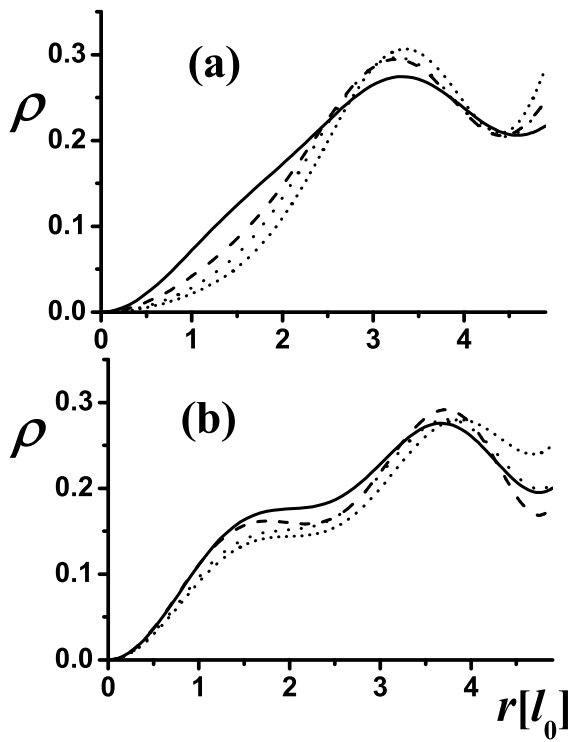}\end{center}
\vspace*{-1cm}
\caption{Electron density distribution in the electron plane when
the neighboring quantum dot contains (a) two electrons and a
hole and (b) three electrons and a hole. The solid line corresponds to
the ground state while the broken lines correspond to the first three
excited states at $M=1,2$ and 3.
}
\end{figure}

In Fig.~3 the electron density distribution in the two-dimensional
electron layer at $\nu=\frac25$ with the QD located at the origin is shown. 
The dot contains either two electrons and one hole (Fig.~3a), or three electrons
and a hole (Fig.~3b). The separations between the electron layer and the 
quantum dot are $a=2.0\, l_0$ and $a=2.25\, l_0$ in Fig.~3a and Fig.~3b, 
respectively. These data correspond to the states presented in Fig.~1d and 
Fig.~2d respectively.  The charge distribution is shown for the unpolarized 
gound state (solid line) and for the first three excited states at $M=1$, 2, 
and 3 (broken lines). These excited states are partially polarized.
Compared to Fig.~3a, the charge distribution for the system with 
three electrons in the quantum dot (Fig.~3b) has a plateau-like 
structure at a distance $r\approx 2 l_0$ from the center of the quantum 
dot. This structure is a direct manifestation of the effectively 
larger size of the quantum dot, as compared to that in Fig.~3a.  
For the fully spin polarized electron fluid Coulomb coupled to
a singly-charged quantum dot \cite{schuller,qhconfine}, or to a 
singly-charged impurity \cite{impure}, the fractionally-cherged quasihole
generated by the ionization process as discussed earlier in the literature
\cite{qhconfine,impure}, {\it moves away} from the QD, located
at the origin, as $M$ is increased. This was indicated by the 
minimum in the charge density being shifted to large $r$ for
increasing $M$. It should be pointed out that the position of the local
minimum at different angular momenta of the charge density correspond
to the orbit radius of the quasihole. In the present case of spin-reversed
ground state and the spin-reversed excitations, the minima in the charge
densities remain {\em near the same position} for all values of $M$. This behavior 
of the charge densities could perhaps be a consequence of the {\it spin-reversed 
quasiholes} \cite{spin_quasihole} being localized entirely due to corrlations. 
We have also calculated the quasihole creation energy at $\nu=\frac25$
\cite{schuller} to be $\sim 0.20$ meV, which can also be measured in exciton 
spectroscopy, as reported earlier for $\nu=\frac13$ \cite{schuller}. PL 
experiments to detect the localized $e/5$ charged spin-reversed quasihole 
would be very exciting.

PL experiments in the FQH regime reported as yet, can be
roughly divided into three main categories: (i) remotely acceptor-doped
heterojunctions \cite{kukush}, (ii) single heterojunctions \cite{turber},
and (iii) single quantum wells \cite{schuller,yusa}. In (i) and (ii), 
influence of the photoexcited valence band holes on the electron liquid
is mostly neglected since the hole is expected to be located at a relatively
large distance $a$ from the electron layer. The situation is different
for the case (iii). There, in a quantum well $a$ is at the maximum of the quantum 
well width which is typically 10-25 nm. The Coulomb interaction
between the electron liquid and the hole in the valence band then plays a
crucial role, and at high magnetic fields, charged excitons can form.
These are bound states of two electrons plus a hole. In high quality
samples, the charged excitons move as quasi free particles in the 2DEG, and 
the strong Coulomb interaction between electrons and holes hinders the
formation of fractionally-charged objects at fractional filling factors
\cite{schuller,yusa}. There are also theoretical 
predictions that at a separation of electrons and  valence band holes
larger than $l_0$, fractionally charged excitons should form \cite{Wojs}.
However, these objects have so far not been detected experimentally.
In our previous studies on quantum wells with {\it moderate}
mobility, we found that under specific conditions, i.e., intermediate
electron density and very low temperature, an electron liquid and 
{\em localized} charged exciton can coexist \cite{schuller}. The
Coulomb coupling of the localized charged exciton to the electron
liquid at $\nu=\frac13$ leads to a reduction in the ground state
energy of the coupled system, which was manifested as an energetic 
anomaly in the experiment. From Fig.~1 we can see that also at $\nu=\frac25$
we would expect a lowering of the ground state energy of the coupled
system as compared to the uncoupled liquid (cf. inset of Fig.~1d).
The magnetic field range where the anomaly appears in Fig.~2 of Ref.~
\cite{schuller} is relatively broad. In this experiment, $\nu=\frac25$ would
be roughly at $B=9$ T, which is just at the onset of the anomaly.
Therefore, it is possible that already in this experiment, there is a 
signature of the spin-reversed quasihole at $\nu=\frac25$ present.
To study this in more detail, it would be highly desirable to have
a better control over the distance, $a$, between the localized charged
excitons and the electron liquid. We propose therefore experiments employing
a special sample structure. The sample should consist of a
layer of self-organized InGaAs quantum dots and a GaAs quantum well,
containing a high mobility electron system which is separated from the
QD layer by a distance $a$. In such a structure, the distance between
the charged dots and the electron liquid can be very precisely 
controlled by the vertical growth of the sample. The QD can be charged
with single electrons via the external gates. Under these conditions, 
with a well defined distance $a$, we would expect much sharper anomalies
in the photoluminescence recombination energies of excitons in the quantum
dots at fractional fillings. Details will be published elsewhere.

In summary, we have investigated the ground state and low-energy
excitations of a quantum Hall liquid at a filling factor $\nu=\frac25$
which is Coulomb coupled to a quantum dot (a liquid-qd complex) separated 
from the liquid by a distance $a$. The dot consists of two or three
electrons and a hole. It was found that the presence of the dot changes locally 
the spin polarization of the electron liquid that depends crucially on the
separation distance. In addition, the electron density distribution in the
quantum Hall layer indicates creation of a quasihole that seems to have
been localized in the electron plane due to the close proximity of the
quantum dot containing interacting electrons. The energy required to create
the quasihole has been calculated. We have also outlined a proposal to detect 
the quasihole properties at $\nu=\frac25$ via PL experiments.

The work of T.C. has been supported by the Canada Research Chair
Program and the Canadian Foundation for Innovation (CFI) Grant.
C.S. would like to acknowledge the support from DFG grant No. SCHU
1171/1-3 and GrK ``Nonlinearity and Nonequilibrium in Condensed
Matter''.

\end{document}